UDK: 538.91

# SIMULATION STUDY OF SHORT CHANNEL EFFECTS IN JUNCTIONLESS SOI MOSFETS

**Khalilloyev Mahkam Muhammadsharifovich**
Lecturer, Physics department
of Urgench State University,
E-mail: x-mahkam@mail.ru

**Annotasiya.** Korilayotgan ishda planar va otishsiz vertical uch zatvorli metal-oksid-yarimotkazgich tranzistorlarda kiska kanalli effect - DIBL effect va transistor otish xarakteristikasi bosagadan past soxadagi kiyaligi SS ga zatvorning yon tomonga kengayishi, kanal kalinligi va legirlanish darajasining ta'sirlari takkoslangan. Natijalar shuni korsatadiki korilgan legirlanish darajasida va kanal kalinligida otishsiz tranzistorda DIBL effect nisbatan kichik kimatlarga ega.

**Kalit so'zlar:** O'tishsiz MOY transistor, qisqa kanal effektlari, stok tomonidan potensial to'siqni kamayishi (DIBL)

**Аннотация.** В рассматриваемой работе сравниваются короткоканльный эффект-DIBL эффект и наклон переходной подпороговой вольтамперной характеристики планарного и вертикального трехзатворного безпереходного транзистора металл-оксид-полупроводник при различном боковом расширении затвора, уровне легирования и толщине канала. Показано, что в рассмотренном диапазоне уровней легирования и толщинах канала DIBL эффект сравнительно ниже для безпереходного транзистора.

**Ключевые слова:** Без перехода МОПТ, короткоканальные эффекты, оксидно-Слив-индуцированного барьера снижение (DIBL)

**Abstract.** In this work influence of gate extension, channel doping level and channel thickness to short channel effects- DIBL effect and subthreshold swing,SS for the planar and vertical junctionless field effect transistors is compared. It is shown in considered range of doping level and channel thickness the DIBL effect is less for junctionless vertical field effect transistor.

**Key words:** Junctionless MOSFET, short-channel effects, Drain-induced barrier lowering (DIBL), SS

**Introduction:** In very short-channel MOSFET devices (L=10 nm or less) the formation of ultrasharp source and drain junctions imposes orders of magnitude of variation in doping concentration over a distance of a few nanometers. Such concentration gradients impose drastic conditions on doping techniques and thermal budget. To avoid this conditions, at last few years it was proposed junctionless MOSFETs [1,2]. The proposed devices are fabricated without the need for forming junctions. Since the channel doping concentration and type are the same as in the source and drain extensions, there is no doping concentration gradient and therefore no impurity diffusion during thermal processing steps. This relaxes the thermal budget by a great deal.

The electrical characteristics of JLMOSFETs are identical to those of normal MOSFETs, but the physics is quite different [3-7]. The heavily doped junctionless transistor is fully depleted below threshold. As gate voltage is increased, the electron concentration n in the channel increases, and threshold is reached when the peak electron concentration in the channel reaches the doping concentration $N_d$. Further increasing the gate voltage increases the ''diameter'' of the region where $n = N_d$, until the entire cross section of the device becomes neutral (i.e. no longer depleted, even partially), at which point flat band is reached.

The primary concern for MOSFET scaling has been the control and suppression of short channel effects (SCEs), such as drain induced barrier lowering (DIBL), degradation of subthreshold slope (SS), that tend to degrade device performance. The use of multiple-gate topologies significantly enhances the electrostatic integrity of the device and provides increased immunity from SCEs [4-9]. To compare the performance of planar and multiple-gate JLMOSFET, in this work DIBL effect and subthreshold slope for planar (in 2D simulation) and trigate (in 3D simulation) SOI Junctionless MOSFET with gate length of 10 nm have been investigated.

In many practical cases for Integrated Circuits (MOSFET memory, CMOS based logic gates) the arranged line of MOSFETs (Fig 2) is used. In this cases all MOSFETs in line can be covered by common gate. This leads to the extended gate of MOSFETs in the line compared to gate contact of the single MOSFET (Fig 3, a). Extended gate results in changing of parasitic capacitances and as a consequence to change short channel effects in nanometer

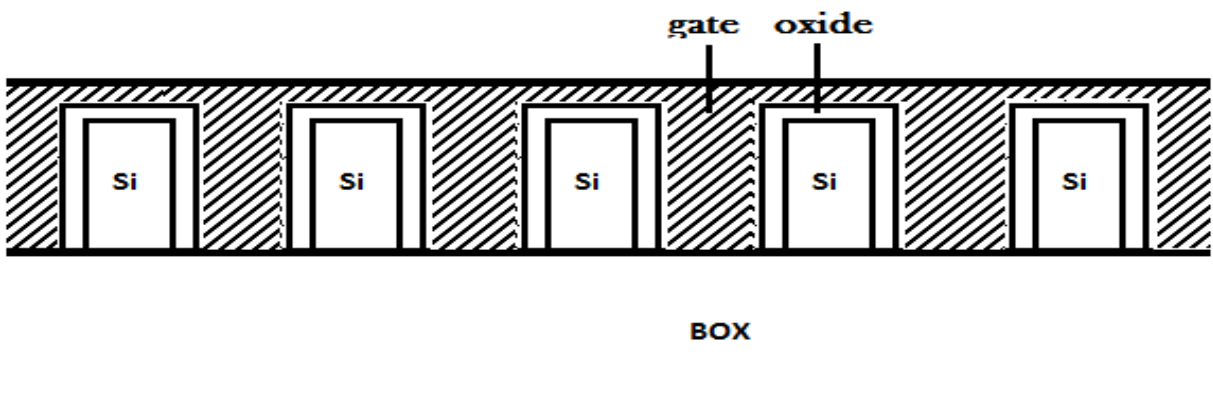

Fig 2. SOI JLMOSFETs in the line

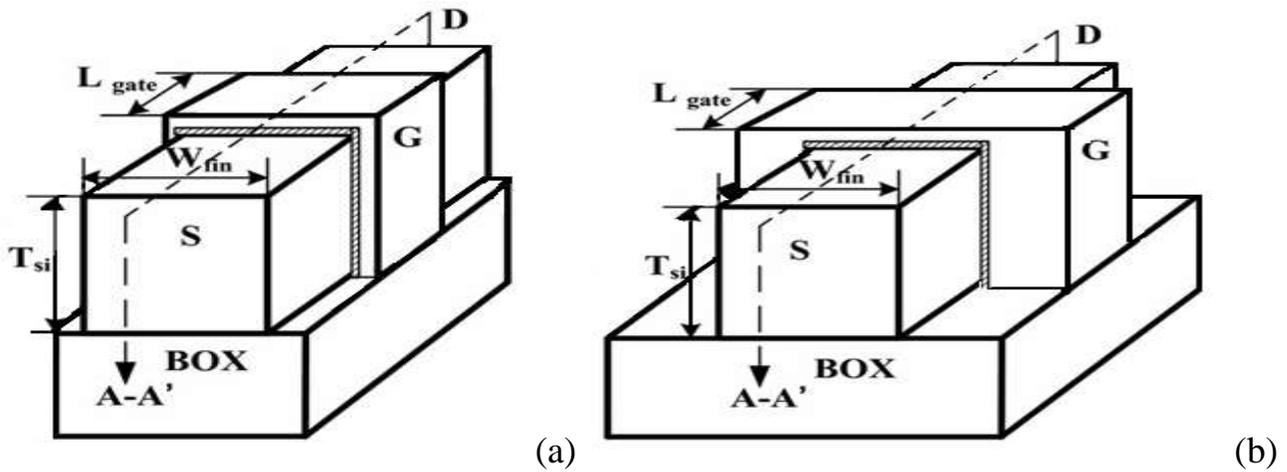

Fig 3. Trigate SOI JLMOSFET with not extended (a) and extended (b) gate.

MOSFETs. For estimation of the influences of the extension of the gate lateral part to short channel effects DIBL effect and subthreshold slope (SS) in trigate SOI JLMOSFET with extended and not extended gates have been compared. For simulations Advanced TCAD Sentaurus device simulator has been used [5-16].

**Results of simulation and discussion.**

We simulated planar SOI JLMOSFET with gate length of 10 nm in 2D (Fig.4) and trigate SOI JLMOSFET with gate length and width of 10nm in 3D (Fig.3 a,b). In the case of trigate JLMOSFET we considered transistors with not extended (Fig 3,a) and extended (Fig 3,b) gates. The thicknesses of the $HfO_2$ gate insulator used in simulation was 0.55 nm.

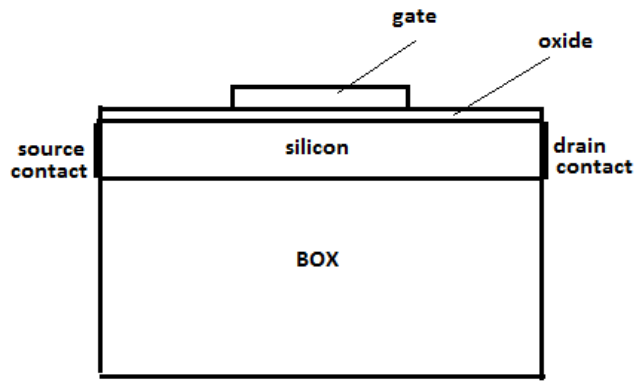

*Fig. 4. Planar SOI JLMOSFET*

For calculation of DIBL and subthreshold slope we simulated transfer characteristics of JLMOSFETs with donor concentration in silicon layer $5·10^{18}$ cm$^{-3}$, $1·10^{19}$ cm$^{-3}$, $5·10^{19}$ cm$^{-3}$ and thicknesses of the active layers Tsi in range 4-10 nm, width of silicon is 10 nm. DIBL is calculated as change of threshold voltage per 1 V changing of drain voltage. The transfer characteristics were simulated for Vds = 0.05V and 0.75 V. In Fig.5 the transfer characteristics for planar 2D and trigate 3D JLMOSFETs with extended and not extended lateral part of gate in case of donor concentrations $5·10^{18}$ cm$^{-3}$ and $5·10^{19}$ cm$^{-3}$ with silicon thickness 4 nm can be seen.

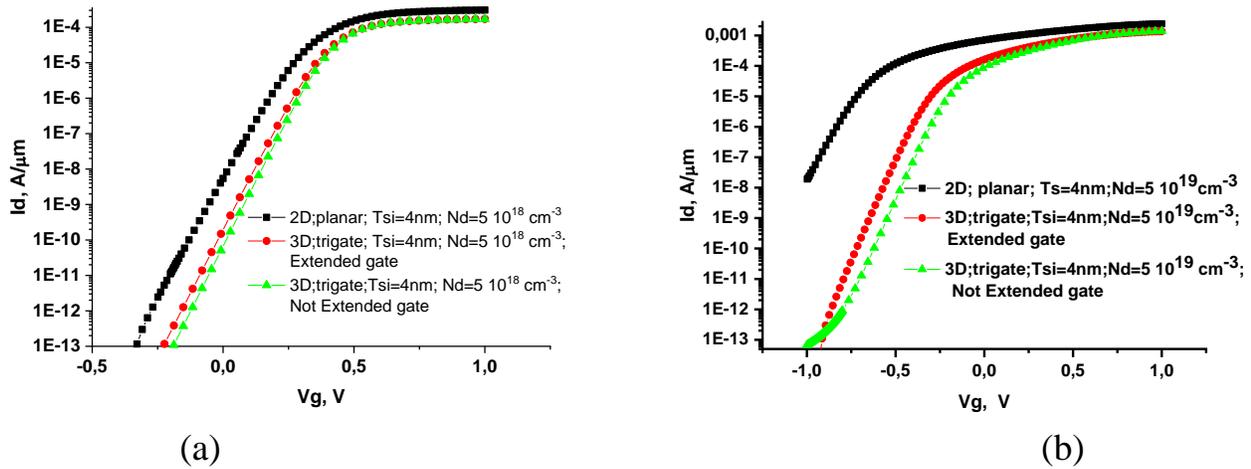

(a)　　　　　　　　　　　　　　　　　　(b)

*Fig.5. Transfer characteristics for planar 2D and trigate 3D JLMOSFETs with extended and not extended lateral part of gate in case of donor concentrations Nd=5·10$^{18}$ cm$^{-3}$ (a) and Nd= 5·10$^{19}$ cm$^{-3}$ (b) with silicon thickness Tsi=4 nm. Vds = 0.75 V*

In comparing of Fig 5 (a) and Fig 5 (b) we can see that for devices with extended and not extended gate the threshold voltage difference is increased with increasing of doping concentration. Lateral extending of gate also leads to more changing of threshold voltage for transistors with more thickness (Fig 6).

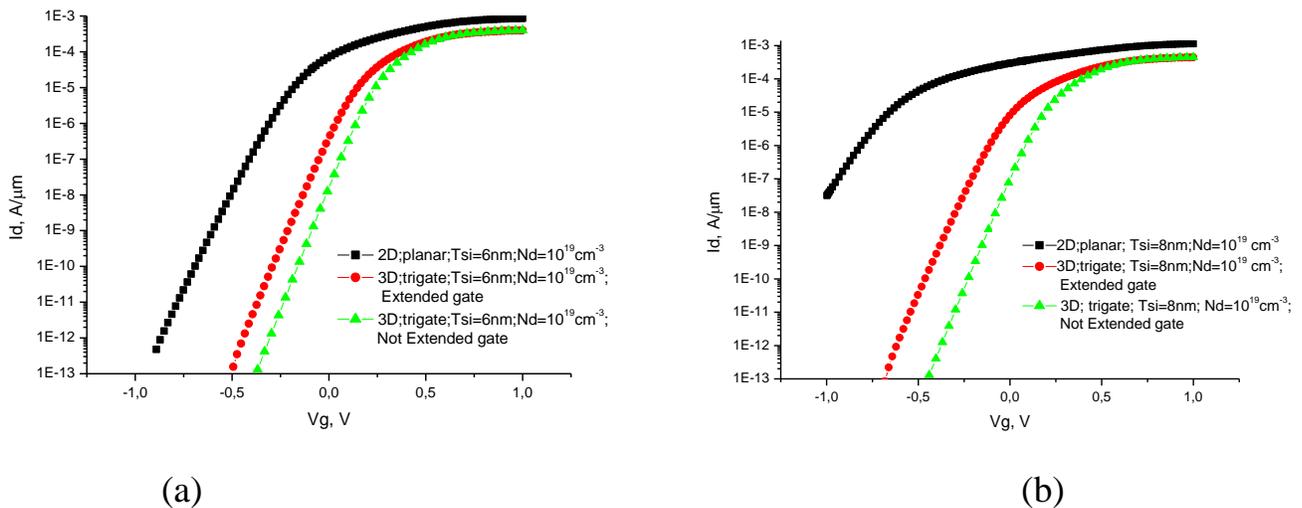

(a)　　　　　　　　　　　　　　　　　　(b)

*Fig.6. Transfer characteristics for planar 2D and trigate 3D JLMOSFETs with extended and not extended lateral part of gate in case of donor concentrations Nd= 5·10$^{19}$ cm$^{-3}$ with silicon thickness Tsi= 6 nm (a) and Tsi=8nm (b). Vds = 0.75 V*

Simulation results show, that short channel effects such as DIBL effect are different for planar and trigate SOI JLMOSFET. DIBL effect is higher for planar than trigate devices in all considered range of silicon thickness (Fig 7) and doping concentrations (Fig 8), and it is monotonously increased with increasing of thickness and doping concentration for planar as well as for trigate JLMOSFET. At high silicon thickness DIBL is saturated for trigate transistor with extended gate, and at Tsi=10nm it is lower than for device with not extended gate (Fig 7 ). Difference between DIBL for planar and trigate devices is increased with increasing of silicon doping concentration (Fig 8).

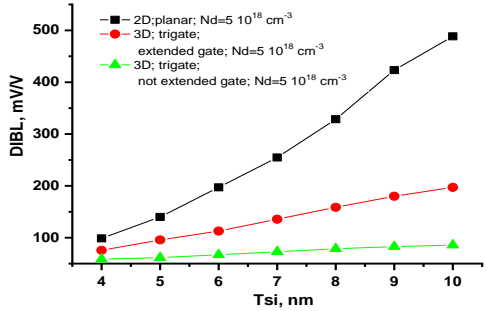
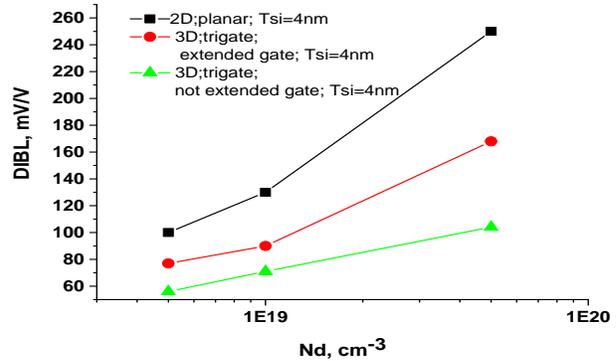

*Fig.7 DIBL dependence on silicon thickness for planar (2D) and trigate (3D) devices with silicon doping concentration Nd=5· $10^{18}$ cm$^{-3}$*

*Fig.8. DIBL dependence on silicon doping concentration for planar (2D) and trigate (3D) devices with silicon thickness Tsi= 4nm.*

Average subthreshold slope (SS) as well as DIBL at silicon thickness Tsi=4nm have not big differences for planar and trigate transistors (Fig 7, Fig 9) while with increasing of silicon thickness and doping concentrations it is increased (Fig 9, Fig 10). Average SS is saturated with increasing of silicon thickness for trigate transistors, while for planar device it growth monotonously (Fig 9). For trigate JLMOSFET with extended gate average SS approximately is same, that of planar device at lower silicon thicknesses up to Tsi=6 nm.

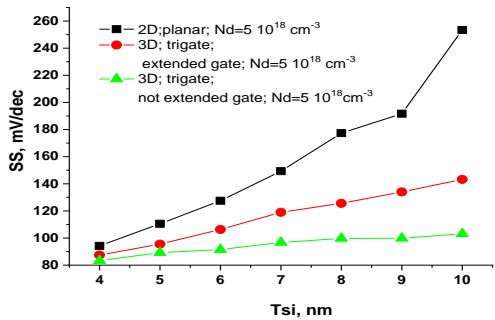
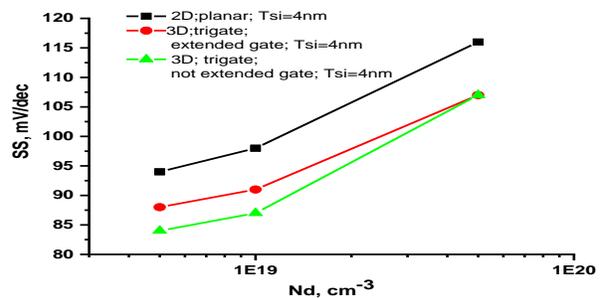

*Fig 9. Average subthreshold slope dependence on silicon thickness for planar (2D) and trigate (3D) transistors with doping concentration Nd= 5·$10^{18}$ cm$^{-3}$ .*
*Fig 10. Average subthreshold slope dependence on doping concentration for planar (2D) and trigate (3D) transistors with silicon thickness Tsi=4nm .*

## Conclusion

Simulation results of SOI JLMOSFET with gate length of 10 nm show that:

1. In all ranges of considered silicon thicknesses and doping concentrations DIBL for trigate devices is lower than for planar devices

2. The average SS for planar and trigate devices with gate length of 10 nm, width of 10nm, silicon thickness of 4 nm and silicon doping concentration of $5 \cdot 10^{18}$ cm$^{-3}$ is in range between 80-100 mV/ dec .

3. DIBL for trigate devices with extended gate is saturated with increasing of silicon thickness and it is higher than for devices with not extended gate up to silicon thickness 8nm.